\documentclass[12pt,a4paper]{article}
\usepackage{epsfig}
\usepackage{graphics}
\usepackage{amsmath}
\newcommand{\bmath}[1]{\mbox{\boldmath{${#1}$}}}

\newcommand{\half}{\mbox{${\textstyle \frac{1}{2}}$}}           % 1/2
\newcommand{\rd}{\textrm{d}}

\begin{document}
%\pagenumbering{}
\setlength{\unitlength}{1mm}
%\def\theequation{\thesection.\arabic{equation}}
%%%%%%%%%%%%%%%%%%%%%%%%%%
%
%------------------------
% page 0
%------------------------
\newpage
\hspace{15cm}

%%%%%%%%%%%%%%%%%%%%%%%%%%%%%%%%%%%%%%%%%%%%%%%%%%%%%%%%%%%%%%%%%
\begin{flushright}
{\bf TSL/ISV-2006-0297 \\
December  2006}
\end{flushright}

\vspace{5mm}

\begin{center}

\vspace*{10mm}

{\Large Pion polarisabilities and bremsstrahlung }
\\[10ex]
G{\"o}ran F{\"a}ldt\footnote{E-mail: faldt@tsl.uu.se}  
\\[3ex]
 Department of nuclear and particle physics\\
Uppsala University\\
 Box 535, S-751 21 Uppsala \\[8ex]

\vspace{5mm}

{\bf Abstract}
\end{center}
  A model for high-energy, small-angle pion-nucleus
bremsstrahlung, $
\pi^- +A\rightarrow\pi^- +\gamma +A$,
is developed within the Glauber diffraction
theory. Special attention is focussed on the
possibility of measuring the pion polarisability in such reactions. 
That is 
the case under the Coulomb peak provided the bremsstrahlung
photon carries practically all the energy of the incident pion.
Only radiation from external legs is considered.

\vspace{5mm}
\noindent

\vfill

\newpage
%
%
%%%%%%%%%%%%%%%%%%%%%%%%%%%
%
\section{Introduction}

Consider a high-energy coherent nuclear production process
\[
	a+A\rightarrow a^{\star}+A 
\]
with quantum numbers exchanged those of the photon. Such a reaction 
can be initiated by a one-photon
exhange, Coulomb production, or by strong interactions. The characteristic
feature of Coulomb production is a very sharp peak near the forward direction.
As energy increases the peak becomes sharper and at the same time moves
towards smaller angles. As a consequence, it becomes possible to disentangle Coulomb
and strong production. If the final state hadron $a^{\star}$ 
is a resonance  the decay rate $\Gamma (a^{\star}\rightarrow a\gamma)$
can be determined. This was first noticed by
Primakoff \cite{1}, and many radiative decay rates have been determined this way.
 A unified theoretical description of both strong and Coulomb production
within the Glauber model was presented in  ref.~\cite{2}. 

The final state hadron $a^{\star}$ need not be a resonance. It can 
also be a multiparticle state. In that case Coulomb production gives information 
on the cross section for the reaction 
$ \gamma + a \rightarrow \gamma +a^{\star}$ \cite{3}. An example is pionic Coulomb 
production or bremsstrahlung,
\[
\pi^- +A\rightarrow\pi^- +\gamma +A
\]
which is driven by low-energy Compton scattering $\gamma+\pi^- \rightarrow \gamma+\pi^-$.
It has been suggested \cite{4}, that by studying pionic Coulomb
production  important information on the 
Compton amplitude can be extracted. The low-energy Compton scattering amplitude
is a sum of two contributions,  a structure-independent 
Thomson term, and a  structure-dependent Rayleigh term. The latter is fixed by the pion
polarisability, and could, under ideal circumstancies be determined
in high-energy nuclear Coulomb production. 
 An experiment with this aim 
 has been performed  \cite{5}, and a reasonable value for the pion 
 polarisability was extracted.

We shall investigate pionic Coulomb production within the Glauber model
\cite{6}. In particular, we are interested in determining the nuclear 
form factors that accompany the various terms in the underlying
Compton amplitude.

Our presentation is arranged as follows. First, we recall the
theoretical description of low-energy pion Compton scattering.
Then, we use this information to develop, in the Born approximation,
the nuclear small-angle amplitude for pionic Coulomb production.
Finally, we show how these expressions are changed when
nuclear multiple scattering is taken into account. Coulomb as well
as nuclear multiple scattering is considered.

%
%
%%%%%%%%%%%%%%%%%%%%%%%%%%%
%
\newpage
\section{Pion Compton scattering} 

The driving amplitude for pionic Coulomb production is the the low-energy
pion Compton amplitude. It is important to understand the origin and structure
of the various contributions to this amplitude since different sub amplitudes
might lead to different form factors when embedded in the nucleus. That
we need the low-energy amplitude can be understood by looking at the the
reaction from a coordinate system where the initial pion is at rest and the 
nucleus runs past at high speed. Since we are in the Coulomb region 
the momentum transfer is extremely tiny and the kick to the pion
consequently very gentle. 
 
The Compton amplitude can be written as
\[
{\mathcal M}(\gamma(q_1)\pi^-(p_1)\rightarrow \gamma(q_2)\pi^-(p_2))=
  {\mathcal M}_{\mu\nu}\epsilon_1^{\mu}(q_1)\epsilon_2^{\nu}(q_2)\ .
\]
Gauge invariance requires that, for real as well as virtual photons
with  $q^2\neq0$, the Compton tensor satisfies 
\[
  {\mathcal M}_{\mu\nu}q_1^{\mu}= {\mathcal M}_{\mu\nu}q_2^{\nu}=0 \ .
\]
We need the dominant contributions to the amplitude at small c.m. 
energies. They are the Born terms and the polarisability terms. 

For pions there are three Born amplitudes described by the 
Feynman diagrams of fig.~1.
\begin{figure}[ht]
\begin{tabular}{c@{}c@{}c@{}}
\scalebox{1.0}{\includegraphics{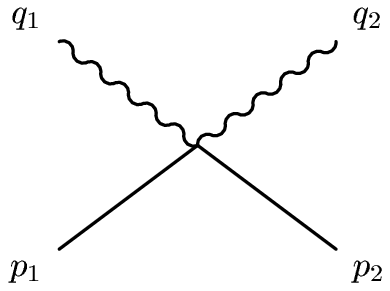}}&\quad
\scalebox{1.0}{\includegraphics{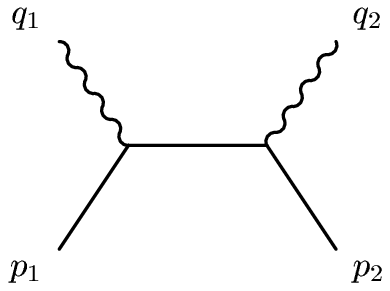}}&\quad
\scalebox{1.0}{\includegraphics{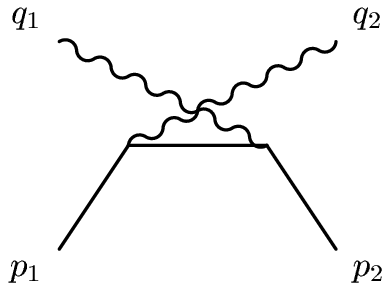}}  
 \end{tabular}
 \caption{Born diagrams for pion Compton scattering.}
  \label{fig:1}
\end{figure}\\
\noindent
Their sum is gauge invariant 
\begin{equation}
  {\mathcal M}_{\mu\nu}^{B}= ie^2\left[ 2g_{\mu\nu}
  -\frac{(2p_2+q_2)_{\nu}(2p_1+q_1)_{\mu}}{2p_2\cdot q_2}
  -\frac{(2p_1-q_2)_{\nu}(2p_2-q_1)_{\mu}}{-2p_1\cdot q_2}\right].
\label{Compton_Born}
\end{equation}
This expression is correct also when $q_1^2\neq 0$, and
that will be important for our applications.   
In the low-energy limit  the amplitude $ {\mathcal M}_{\mu\nu}^{B}$
 reduces to the structure-independent Thompson term. 

Terms quadratic in the photon energies are structure dependent,
Rayleigh terms.  They come from diagrams like
\begin{figure}[ht]
\begin{tabular}{c@{}c@{}c@{}}
\scalebox{1.10}{\includegraphics{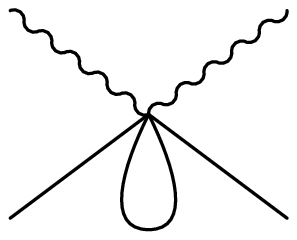}}&\quad
%\scalebox{1.10}{\includegraphics{ray2.eps}}&\quad
\scalebox{1.10}{\includegraphics{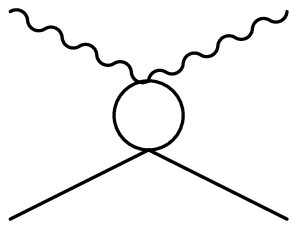}} &\quad
 \scalebox{1.10}{\includegraphics{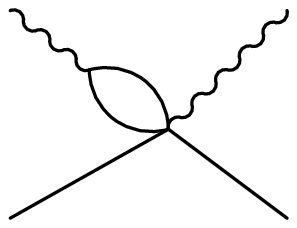} }
 \end{tabular}
  \caption{One-loop Feynman diagrams in chiral-Lagrangian theory.}
\end{figure}\\
\noindent
In second order of the photon momenta we can construct 
two gauge invariant Compton tensors, which we choose to be
\begin{equation}
 {\mathcal M}_{\mu\nu}^{I}= ie^2\,
        \displaystyle{       \frac{2m_{\pi} \beta_{\pi}}{\alpha} }
   \bigg[ q_1\cdot q_2\, g_{\mu\nu} -q_{2\mu} q_{1\nu} \bigg]  
\label{Compton_pol_I}
\end{equation}  
with $\beta_{\pi}$ the magnetic polarisability, and
\begin{equation}
\begin{array}{rl}
 {\mathcal M}_{\mu\nu}^{II}=& ie^2  \,
   \displaystyle{ \frac{-2(\alpha_{\pi}+\beta_{\pi})}{m_{\pi}\alpha} }
   \bigg[ p_1\cdot q_1 p_1\cdot q_2\, g_{\mu\nu}
        +q_1\cdot q_2\, p_{1\mu}p_{1\nu} \\
 &  -p_1\cdot q_1 \, q_{2\mu}p_{1\nu} 
   -p_1\cdot q_2\,  p_{1\mu}q_{1\nu} \bigg] 
\end{array}
\label{Compton_pol_II}
\end{equation}  
with $\alpha_{\pi}$ the electric polarisability.
The  Compton amplitude  is the sum of the above individual 
contributions, expressions (\ref{Compton_Born}), (\ref{Compton_pol_I}) 
and (\ref{Compton_pol_II}). 
Evaluated in the  lab system (initial pion at rest) the Compton
amplitude reads
\[
\begin{array}{rl}
 {\mathcal M}=& i8\pi m_{\pi} \big[
  \displaystyle{ -\frac{\alpha}{m_{\pi}}   }
  \bmath{\epsilon}_1\cdot  \bmath{\epsilon}_2 \\
& +  \alpha_{\pi}\omega_1\omega_2 \bmath{\epsilon}_1\cdot  \bmath{\epsilon}_2
  +\beta_{\pi}(\bmath{q}_1\times\bmath{\epsilon}_1)\cdot
       (\bmath{q}_2\times\bmath{\epsilon}_2)\big] \ .
\end{array}
\]

In ChPT one obtains in the one-loop order \cite{7}
\[
 \alpha_{\pi^\pm}=- \beta_{\pi^\pm}=2.7\cdot10^{-4} \ \mbox{\rm fm}^3.
\]
In this order one has,  $\alpha_{\pi^\pm}+ \beta_{\pi^\pm}=0$, and
we shall assume this relation throughout. Also two-loop contributions to 
the polarisabilities have been calculated \cite{8}, and found to be small.
Therefore, we  put ${\mathcal M}_{\mu\nu}^{II}=0.$ It is useful to 
introduce the dimensionless parameter
\begin{equation}
 \lambda=    \frac{\beta_{\pi^\pm}m_{\pi^\pm}^3 }{\alpha }
   = - 0.013 \ .
\label{lambda-def}
\end{equation}
There is no agreement on the experimental value of the polarisability 
so we shall use the value (\ref{lambda-def}) as a guide in our deliberations.  
%
%
%
%%%%%%%%%%%%%%%%%%%%%%
\newpage
\section{Pionic bremsstrahlung: Born approximation}

Pionic bremsstrahlung can be accompanied both by electromagnetic
and strong interactions. We are interested in the region
of small momentum transfers to the nucleus, and in particular the Coulomb region.
This means that when the pion interaction with the nucleus is
through a one-photon exchange, then the relevant Compton
amplitude is a low-energy amplitude, as desired.

The kinematics of pionic bremsstrahlung  is
\[
\pi^-(p_1) +A(p)\rightarrow\pi^-(p_2) +\gamma(q_2) +A(p') \ .
\]
Our interest is focussed on coherent high-energy interactions 
where the energies of the pions, as well as that of the radiated 
photon, are many GeV:s. Coherence demands that the transverse
momenta $\bmath{p}_{2\bot}$ and $\bmath{q}_{2\bot}$ 
be much smaller than the inverse of the nuclear radius $R_A$,
and as a consequence of the high energies, the longitudinal momentum transfer to
the nucleus may be ignored. Thus, to a precision sufficient for the 
subsequent analysis
we put
\[
p_{1z} = p_{2z} +q_{2z} 
\]
with $z$ denoting the longitudinal direction. The energy transfer
to the nucleus can likewise be ignored.

The one-photon exchange graph is pictured in fig.~3. The small blob in the
\begin{figure}[h]\begin{center}
\scalebox{1.20}{\includegraphics{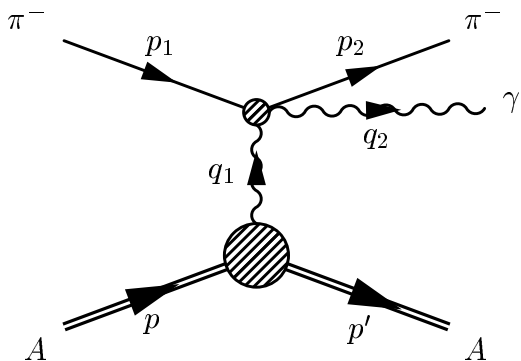}}
\end{center}
\caption{Born diagram for pionic bremsstrahlung.}
\end{figure}
graph represents the full 
pion-Compton amplitude;
the large blob the photon-nucleus electromagnetic vertex.
The pion charge is $-e$, the nuclear charge $Ze$, and 
the nucleus is treated as a spin-zero particle. 
With $q_1$ the virtual photon four-momentum, these
assumptions lead to a
Coulomb production amplitude  
\begin{equation}
  {\mathcal M}_{C}=\frac{-i}{q_1^2} 
   {\mathcal M}_{\mu\nu}(p_2,q_2;p_1,q_1)(-iZe)
             (p+p')^{\mu}\epsilon_2^{\nu} \ .
\label{Coulomb_Born}
\end{equation}
Since the Compton tensor $ {\mathcal M}_{\mu\nu}$ is 
 gauge invariant we may also make the
replacement $p+p'=2p+q_1\rightarrow 2p$.

The expression for  ${\mathcal M}_{C}$  is covariant and valid
in all coordinate systems. We prefer to work in the lab system where the 
initial nucleus, of mass $M_A$, is at rest. Choosing a polarisation 
vector $\epsilon_2$ with 
vanishing time  component, leads to 
\begin{equation}
 \begin{array}{rcl}
  {\mathcal M}_C & =&  8i\pi M_A\, \displaystyle{ \frac{-2Z\alpha}{q_1^2} }\,e
  \bigg[ 
  \left( \frac{E_1\bmath{p}_2\cdot \bmath{\epsilon}_2}{p_2\cdot q_2}
 -\frac{E_2\bmath{p}_1\cdot \bmath{\epsilon}_2}{p_1\cdot q_2}\right)
  \\ && \\ && \qquad  \qquad\qquad \qquad \qquad 
   + \displaystyle{\frac{\lambda \omega_2}{ m_{\pi}^2} }\,
      (\bmath{p}_2-\bmath{p}_1)\cdot \bmath{\epsilon}_2
   \bigg] \ .
\end{array}\label{Coulomb_Born_lab}
\end{equation}

Here, the first two terms  originate from the Compton Born 
terms of eq.(\ref{Compton_Born}).
The $g_{\mu\nu}$ Born term does not  contribute to $ {\mathcal M}_C$.
Instead, the first term  of (\ref{Coulomb_Born_lab}) originates from the middle term
of eq.(\ref{Compton_Born}) and represents a pion-nucleus elastic  scattering step 
followed by a pionic bremsstrahlung step. Similarly,
 the second term of (\ref{Coulomb_Born_lab}) 
originates from the last term of eq.(\ref{Compton_Born}) and represents a
pionic bremsstrahlung step followed by a pion-nucleus elastic scattering step.

The last term of (\ref{Coulomb_Born_lab}), finally, is generated by the polarisability contribution,
eq.(\ref{Compton_pol_I}), to the Compton amplitude. In this term the 
radiated and exchanged photons are attached to the same vertex.
The second polarisability contribution, eq.(\ref{Compton_pol_II}),
vanishes since we presume $\alpha_{\pi}+\beta_{\pi}=0$.

Pionic bremsstrahlung from the external pion legs can also occur 
in nuclear interactions, as depicted in fig.~4. 
For $\pi^-$ scattering,  with  $q_1=p-p'$
\begin{figure}[ht]\begin{center}
\begin{tabular}{c@{}c@{}}
\scalebox{1.40}{\includegraphics{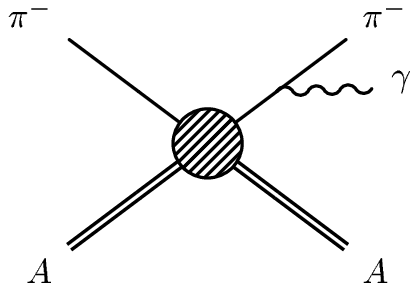}}&\qquad
\scalebox{1.40}{\includegraphics{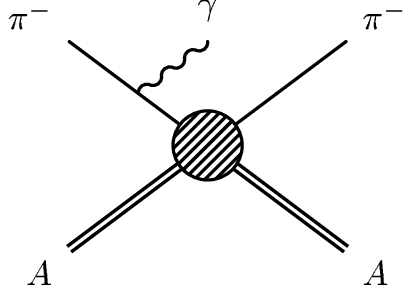}}
 \end{tabular}\end{center}
 \caption{Pion bremsstrahlung in nuclear scattering.}
\end{figure}
the four-momentum transfer to the nucleus, as above, the nuclear
contribution to pionic bremsstrahlung reads
\begin{equation}
  {\mathcal M}_N = e\bigg[  - \frac{\epsilon_2\cdot p_2}{p_2\cdot q_2}
   {\mathcal M}_{\pi A}(s_1,q_1)
  +  \frac{\epsilon_2\cdot p_1}{p_1\cdot q_2}
    {\mathcal M}_{\pi A}(s_2,q_1) 
    \bigg] \ .
\label{Brems-init}
 \end{equation} 
Here, ${\mathcal M}_{\pi A}(s,q)$ is the elastic pion-nucleus scattering amplitude
at energy $s$ and momentum transfer $q$. The nuclear scattering steps occur with the same
momentum transfer. In the first term of (\ref{Brems-init}) the
pion-nucleus scattering occurs after the photon is radiated, and the 
amplitude should  therefore be evaluated at energy $s_2=(p_2+p')^2$.
Similarly,  the pion-nucleus amplitude of the second term of eq.(\ref{Brems-init})
is evaluated at $s_1=(p_1+p)^2$.
To insure gauge invariance of ${\mathcal M}_N$ one could, {\it e.g.},
evaluate the  pion-nucleus amplitudes at
the same energy $s$. The alternative is to add an appropriate 
counter term arising from internal radiation. We shall not do so, however,
but shall take the amplitude (\ref{Brems-init}) as it stands. The error 
committed is probably negligible. 

The pion-nucleus amplitude $ {\mathcal M}_{\pi A}(s,q)$
is related to the pion-nucleus elastic scattering amplitude in the 
lab system, $F_N(E,\bmath{q})$, through
\begin{equation}
   {\mathcal M}_{\pi A}(s,q)=8i\pi M_A \, F_N(E,\bmath{q})
\label{M-lab_amp}
 \end{equation}
 with $E$ the pion lab energy. The  amplitude $F_N(E,\bmath{q})$
 depends only on the transverse part $\bmath{q}_\bot$ of $\bmath{q}$. 

In our application we are in an energy region where the pion-nucleus
total cross section $\sigma_{\pi A}$ may be considered independent of energy. 
If in addition, we specialise to the region of
Coulomb production the dependence on momentum transfer can
be ignored, and  a simple energy dependence emerges
\begin{equation}
   F_N(E,0)=\frac{ik}{4\pi}\sigma_{\pi A}(1-i\alpha_{\pi A})
 \equiv\frac{ik}{4\pi}\sigma_{\pi A}' \ .
 \end{equation}
Since at high energies there is no
distinction between $k$ and $E$
we can write the nuclear bremsstrahlung contribution as
\begin{equation}
  {\mathcal M}_N =8i\pi M_A \,\frac{i e\sigma_{\pi A}'}{4\pi}
  \bigg[ 
  \frac{E_1 \bmath{\epsilon}_2\cdot \bmath{p}_2}{p_2\cdot q_2}
   - \frac{E_2\bmath{\epsilon}_2\cdot \bmath{p}_1}{p_1\cdot q_2}
    \bigg]  \ .
\label{Nuclear_brems}
 \end{equation}
The factor inside the brackets is identical to the corresponding
 factor in eq.(\ref{Coulomb_Born_lab}) describing radiation
 from the external legs. 
 
The strong  pion-nucleus amplitude $ {\mathcal M}_{\pi A}$ 
is conveniently calculated in the
Glauber model. Further details below.
%
%
%
%%%%%%%%%%%%%%%%%%%%%%%%%
\newpage
\section{Elastic Coulomb scattering}

At this point it is useful to recall the eikonal description
of elastic Coulomb scattering. The Coulomb potential
for $\pi^-$-nucleus scattering is
\begin{equation}
  V_C(\bmath{r})= -\frac{Z\alpha}{r}
\label{Coul_pot}
 \end{equation}
and the scattering amplitude in the Born approximation 
\begin{equation}
  f_C(E,\mathbf{q})= -\frac{E}{2\pi}\int\rd^3x\, 
   e^{-i\mathbf{q}\cdot\mathbf{x}}\,V_C(\mathbf{r})
   =\frac{2Z\alpha E}{\mathbf{q}^2}
\label{Rel_Born}
\end{equation}
with the momentum transfer $\bmath{q}=\bmath{k}'-\bmath{k}$. 
In reality, at high energies the momentum tranfer is 
transverse so that $\mathbf{q}=\mathbf{q}_\bot$.

Elastic Coulomb scattering to all orders in the fine-structure constant
is exactly reproduced by the eikonal approximation 
\begin{equation}
  f_C(E,\mathbf{q})= \frac{ik}{2\pi }
    \int\rd^2b\, e^{-i\mathbf{q}\cdot\mathbf{b}}
   [1 - e^{i\chi_C(\mathbf{b})}]
\label{Rel_eik}
\end{equation}
with the point-like Coulomb phase function
\begin{eqnarray}
  \chi_C(\mathbf{b})&=& -\eta \ln (b/2a)
                       \label{Point-Coul-phase}\\
  \eta&=& 2Z\alpha/v \ .
\end{eqnarray}
Here, $a$ is a cut-off parameter introduced in order to 
make the phase function finite. The Coulomb potential is
cut off at a radius $r=a$.

At high energies there is no distinction between $E$ and $k$, and
$v=1$. A straightforward integration gives the well-known result
\begin{eqnarray}
  f_C(E,\mathbf{q}) &=& \frac{2Z\alpha E}{\mathbf{q}^2}
   \exp [i\sigma_{\eta} + i\eta \ln (aq)] \label{Point-cross}\\
    \sigma_{\eta} &=& 2\arg\Gamma(1-i\eta/2) \ .
\end{eqnarray}

A thorough description of Coulomb scattering in the eikonal
approximation can be found in 
\cite{6}, but also \cite{9} contains useful information. For an extended
charge distribution the phase-shift function takes the form
\begin{equation}
	\chi_C(b)=-\eta \left[\ln(b/2a)
	    + 2\pi \int_b^{\infty}T(b')\ln(b'/b)b'\rd b'\right]
	     \label{Coul-ext-ch}
\end{equation}
where the target-thickness function $T(b)$ corresponds to a 
nuclear charge-density distribution
normalised to unity.

In the real world pions obey the Klein-Gordon equation which contains
one term that is linear in the Coulomb potential $V_C(\bmath{r})$, 
and one which is quadratic. The above applies to the linear term. 
The quadratic term leads to corrections that can be expanded 
in powers of $\sin\theta/2=q/2k$. For a point charge in the Klein-Gordon
case \cite{10}
\begin{eqnarray}
	f_C^{KG}(E,\mathbf{q})&=&f_C(E,\mathbf{q})
	 \left\{ 1+\frac{\pi(Z\alpha)^2e^{i\sigma_1}}{\eta}\frac{q}{2k} + \ldots \right\}\\
	 \sigma_1&=& 2 [\arg \Gamma(\half+i\eta/2) -\arg \Gamma(1+i\eta/2) ] \ .
\end{eqnarray}
A few of the higher order corrections have also been evaluated \cite{11}.
The important point for us is that we consider the very small 
angular region where the correction terms safely can be  neglected.
The corrections due to the extension of the nuclear charge 
distribution are negligible as well.

%
%
%
%%%%%%%%%%%%%%%%%%%%%%%%%%%%

\newpage
\section{Pionic bremsstrahlung: eikonal approximation}

The Born amplitudes are strongly modified by nuclear multiple
scattering, a modification most easily calculated in the
eikonal approximation. But first a  remark on the four-momentum 
transfer to the nucleus, $q_1$. Its time component can always be
neglected, so that
\begin{equation}
 - q_1^2=  \mathbf{q}_1^2 = \mathbf{q}_{1\bot}^2+q_{1z}^2 \ .
\end{equation}
The longitudinal momentum transfer $-q_{1z}$ is tiny and can
be replaced by its minimum value 
\begin{equation}
	q_{min}
	  =\frac{m^2\omega_2}{2E_1E_2} \label{min-mom}
\end{equation}
but we shall start by neglecting $q_{1z}$ altogether and later return to
the modifications dictated by its  presence.

In the Coulomb-induced bremsstrahlung amplitude (\ref{Coulomb_Born_lab}),
 Coulomb scattering in the one-photon approximation is 
described by the factor
\begin{equation}
\frac{2Z\alpha E_1}{\mathbf{q}_{1\bot}^2} \ .\label{new-Born}
\end{equation}
The very first term of (\ref{Coulomb_Born_lab}) represents 
Coulomb scattering followed by  brems\-strahlung. To include multiple
Coulomb scattering we add all diagrams with multiple-photon exchanges 
between the pion and the nucleus. If all these exchanges 
occur before the bremsstrahlung step then we simply
replace the Born amplitude (\ref{new-Born}) by the full
Coulomb amplitude of eq.(\ref{Rel_eik}). The contributions
from the diagrams where radiation occurs from an intermediate
pion line, {\it i.e.}, when we have photon exchanges both before and 
after the radiation step, cancel to a large extent. Therefore,
diagrams with radiation from internal lines are ignored. The
corresponding argument applies to the second term of 
(\ref{Coulomb_Born_lab}) as well.

The nuclear-induced bremsstrahlung amplitude (\ref{Brems-init}), 
must be similarly modified.
The pion can radiate when between two nucleon
collisions, if we view the nuclear interaction as caused by
multiple interactions between the pion and the nucleons
of the nucleus. Again, the contributions from the internal
radiation diagrams cancel when summed.

A further generalisation is to include Coulomb and nuclear exchanges
simultaneously. Since internal radiation 
diagrams are ignored we get
\begin{equation}
  {\mathcal M}_{rad}= 8\pi iM_A e  \bigg[ -
  \displaystyle{ \frac{E_2\bmath{p}_1\cdot \bmath{\epsilon}_2}{p_1\cdot q_2} }
     +
  \displaystyle{ \frac{E_1\bmath{p}_2\cdot \bmath{\epsilon}_2}{p_2\cdot q_2}}
    \bigg]{\mathcal F}_{rad}(\bmath{q}_{1\bot}) \ . \label{Ext-Ka}
\end{equation}

The factor $ {\mathcal F}_{rad}(\bmath{q}_\bot)$ is 
the pion-nucleus 
scattering amplitude, including both its Coulomb and nuclear
interactions. In the Glauber model
\begin{equation}
  {\mathcal F}_{rad}(\bmath{q}_\bot)=  
  \frac{iv}{2\pi }
    \int\rd^2b\, e^{-i\mathbf{q}_\bot\cdot\mathbf{b}}
   \left( 1 - e^{i(\chi_C(\mathbf{b})+\chi_N(\mathbf{b}))}\right)
 \label{FGl_C+N}
\end{equation}
with the momentum transfer $\mathbf{q}_\bot$ in the impact parameter plane. 
The function 
${\mathcal F}_{rad}(\bmath{q}_\bot)$ is energy independent, providing 
we can put $v=1$ and neglect the  energy dependence of the pion-nucleus
potentials. The phase-shift functions are related to
the corresponding potentials, whether Coulomb or nuclear,
through
\begin{equation}
  \chi(\mathbf{b})=  
  \frac{-1}{v }
    \int_{-\infty}^{\infty}\rd z\,V(\mathbf{b},z) \ .
 \label{Def-phase-fcn}
\end{equation}

We can, in a well-known fashion, decompose ${\mathcal F}_{rad}(\bmath{q})$ 
into a purely 
Coulomb and a Coulomb-distorted nuclear amplitude by writing
\begin{eqnarray}
	{\mathcal F}_{rad}(\bmath{q}_\bot)&=&  
	    \frac{iv}{2\pi }
    \int\rd^2b\, e^{-i\mathbf{q}_\bot\cdot\mathbf{b}}
   \left( 1 - e^{i\chi_C(\mathbf{b})}\right)\nonumber \\
      &&+
  \frac{iv}{2\pi }
    \int\rd^2b\, e^{-i\mathbf{q}_\bot\cdot\mathbf{b}}e^{i\chi_C(\mathbf{b})}
   \left( 1 - e^{i\chi_N(\mathbf{b})}\right)\nonumber\\
     &\equiv & [f_C(E, \bmath{q}_\bot)+f_{N}(E, \bmath{q}_\bot)]/E \ .
 \label{Decomp-C+N}
\end{eqnarray}
The analytic expression of the Coulomb  amplitude
  $f_C(E, \bmath{q}_\bot)$ for a point charge distribution 
  is  given in eq.(\ref{Point-cross}).

The polarisation-induced bremsstrahlung amplitude is the third
term of (\ref{Coulomb_Born_lab}).  Before proceeding we note  the identities
$(\bmath{p}_2-\bmath{p}_1)\cdot \bmath{\epsilon}_2= 
\bmath{q}_1\cdot \bmath{\epsilon}_2$ and 
\begin{equation}
  \displaystyle{ \frac{2Z\alpha}{\bmath{q}_1^2} } \,
        \bmath{q}_1\cdot \bmath{\epsilon}_2 
  =  \displaystyle{\frac{i}{2\pi}} \int\rd^3r\, 
   e^{-i\mathbf{q}_1\cdot\mathbf{r}}\, \bmath{\epsilon}_2\cdot \bmath{\nabla}
   V_C(\bmath{r}) \ .\label{simp-pol}
\end{equation}
The polarisation potential is hence proportional to the gradient of 
the Coulomb potential. We now replace the plane waves of the Born approximation 
by distorted waves,
 assume energy-independent  pion-nucleus
potentials and neglect  the longitudinal momentum transfer $q_{1z}$.
  Introducing phase-shift functions instead of 
potentials leads to the energy-independent expression,
  $v=v_1=v_2$,
\begin{equation}
  \displaystyle{\frac{-i v}{2\pi}} \int\rd^2b\, 
   e^{-i\mathbf{q}_{1\bot}\cdot\mathbf{b}}\, 
   \bmath{\epsilon}_2\cdot \bmath{\nabla} \chi_C(\mathbf{b})
  e^{ i(\chi_C(\mathbf{b})+\chi_N(\mathbf{b})) } \ .
  \label{Eik_pol-factor}
\end{equation}
With help of the identity
\begin{equation}
  e^{ i(\chi_C(\mathbf{b})+\chi_N(\mathbf{b})) }=
   e^{ i\chi_C(\mathbf{b})} - e^{ i\chi_C(\mathbf{b})}
   (1-e^{ i\chi_N(\mathbf{b})})\label{subdiv}
\end{equation}
we can easily isolate a purely Coulombic term from a rest term
which involves also nuclear effects. 
After  partial integration of the first term of (\ref{subdiv}) 
and an angular integration of the second one, expression
(\ref{Eik_pol-factor}) can be written as
\begin{equation}
	\mathbf{q}_{1\bot}\cdot\bmath{\epsilon}_2{\mathcal F}_{pol}(\bmath{q}_{1\bot}) \ .
\end{equation}
The energy-independent form factor ${\mathcal F}_{pol}(\bmath{q}_{1\bot})$ is
\begin{equation}
	{\mathcal F}_{pol}(\bmath{q}_{\bot})=
	[f_C(E,\mathbf{q}_\bot) + f_{P}(E,\mathbf{q}_\bot) ]/E
\end{equation}
with $f_C(E,\mathbf{q}_\bot)$ the elastic Coulomb scattering amplitude 
and the new amplitude $f_{P}(E,\mathbf{q}_\bot)$
 defined by, $q_\bot=\| \mathbf{q}_\bot\|$, 
\begin{equation}
	f_{P}(E,\mathbf{q}_\bot)=k\int_0^{\infty}b^2\rd b\, \frac{J_1(q_\bot b)}{q_\bot b}
	  \frac{\rd \chi_C(b)}{\rd b}e^{ i\chi_C(b)}(1-
  e^{i \chi_N(b)}) \ .\label{f-pol-fcn}
\end{equation}
The last factor of the integrand guarantees that the integrand 
vanishes outside
the nuclear mass distribution.
From the definition of the Coulomb phase for extended
charge distributions, eq.(\ref{Coul-ext-ch}), we derive
\begin{equation}
  \frac{\rd \chi_C(b)}{\rd  b}= -2\pi\eta \frac{1}{b}
   \int_0^b b'\rd b'\, T(b') \ .
\end{equation}

The result of all this is a polarisation contribution to the bremsstrahlung amplitude
\begin{equation}
  {\mathcal M}_{pol}=  8\pi i e M_A \bigg[ \frac{ \lambda}{m_{\pi}^2}   \, \omega_2 
   \mathbf{q}_1\cdot \bmath{\epsilon}_2
    \bigg]{\mathcal F}_{pol}(\bmath{q}_{1\bot})\ .\label{Pol-Ka}
\end{equation}
To get  the complete  ${\mathcal M}$-amplitude we add the amplitude 
${\mathcal M}_{rad}$ representing radiation from external legs;
\begin{eqnarray}
  {\mathcal M}&= & 8\pi i e M_A \bigg\{ \big[ -
  \displaystyle{ \frac{E_2\bmath{p}_1\cdot \bmath{\epsilon}_2}{p_1\cdot q_2} }
     +
  \displaystyle{ \frac{E_1\bmath{p}_2\cdot \bmath{\epsilon}_2}{p_2\cdot q_2}}
    \big]{\mathcal F}_{rad}(\bmath{q}_{1\bot}) \nonumber \\
   && \quad \quad +
  \big[ \frac{ \lambda}{m_{\pi}^2}   \, \omega_2 
   \mathbf{q}_1\cdot \bmath{\epsilon}_2
    \big]{\mathcal F}_{pol}(\bmath{q}_{1\bot}) \bigg\}\ . \label{Full-M-amp}
\end{eqnarray}

So far we have kept the cut-off parameter $a$ of the Coulomb potential. This
was intentional. We now remove the dependence on this parameter by 
removing a phase factor common to all amplitudes. This is 
conventionally done so that the Coulomb phase in the integrand of the 
nuclear amplitude,
$f_{N}(E,\bmath{q}_\bot)$ eq.(\ref{Decomp-C+N}), varies as slowly as possibly
over the domain of integration. Since the main
contribution  comes from a region close to the nuclear rim, 
this goal is achieved by setting $2a=R_u$ in eq.(\ref{Coul-ext-ch}). Here, $R_u$
is the equivalent radius of the uniform nuclear mass distribution.

Let us now return to the question of the longitudinal momentum 
transfer, $q_{min}$. In the nuclear and polarisation amplitudes,
eqs (\ref{Decomp-C+N}) and (\ref{f-pol-fcn}), the integration is 
over the nuclear region, which has a finite extension. Since 
$q_{min}R_u\ll 1$ all dependence on $q_{min}$ can be ignored. 
Extracting  appropriate phase factors we may write, 
with $q_\bot=\|\mathbf{q}_\bot\|$,
\begin{eqnarray}
	f_{N}(E,\mathbf{q}_\bot) &= &
     ik e^{i\delta_B} \int b\rd b\, J_0(q_\bot b)
   \left( 1 - e^{i\chi_N(b)}\right)
 \label{fCN-new} \\
 f_{P}(E,\mathbf{q}_\bot)& =&
	    k e^{i\delta_{B}'}
    \int_0^{\infty}b^2\rd b\, \frac{J_1(q_\bot b)}{q_\bot b}
	  \frac{\rd \chi_C(b)}{\rd b}
	  (1-e^{i \chi_N(b)})
 \label{fCP-new}
\end{eqnarray}
for the nuclear and polarisation amplitudes. The Bethe phases $\delta_B$
and $\delta_{B}'$ need not be factorised as indicated. The original
integrations could also have been carried out.

The pure Coulomb amplitude is more of a problem. In the Born approximation,
as layed out in eq.(\ref{Coulomb_Born}), the Coulomb denominator 
contains the longitudinal momentum transfer through 
$ \mathbf{q}_1^2 = \mathbf{q}_{1\bot}^2+q_{min}^2 $. When $q_\bot\gg q_{min}$
the dependence on $q_{min}$ can safely be ignored, but under the Coulomb peak,
where $q_\bot\approx q_{min}$ that is not possible. A subsequent question 
is then how the nuclear form factor depends on $q_{min}$. This 
question has not yet been resolved analytically. Of course, the scattering
amplitude can be calculated numerically, but until we have done so we
suggest the point-charge Coulomb amplitude be chosen as 
\begin{equation}
  f_C(E,\mathbf{q}_\bot) = \frac{2Z\alpha E}{\mathbf{q}_\bot^2 +q_{min}^2}\, 
   \exp [i\sigma_{\eta} + i\eta \ln (q_\bot R_u/2)] \ .\label{Point-cross-final}
\end{equation}
The form factor for extended charge distributions can, 
as is necessary for large momentum transfers, be calculated as in ref.\cite{9}.
\newpage
\section{Cross sections}

The cross section distribution in the lab system is
\begin{equation}
	\rd \sigma =\frac{1}{4p_1 M_A} \left| {\mathcal M}\right|^2 {\rm dLips}
\end{equation}
where $p_1$ is the incident pion lab momentum. The Lorentz-invariant phase space
is as always
\begin{eqnarray}
{\rm dLips} &=& (2\pi)^4 \delta(p_1+p_A-p_2-q_2-p_A')
  \frac{\rd^3p_A'}{ (2\pi)^3 2E_A'}  \frac{\rd^3p_2}{ (2\pi)^3 2E_2 }
  \frac{\rd^3q_2}{ (2\pi)^3 2\omega_2} \nonumber \\
  &=& \frac{1}{16\pi M_A} \frac{\rd^2 p_{2\bot}}{(2\pi)^2}
    \frac{\rd^2q_{2\bot}}{(2\pi)^2}\frac{\rd q_{2z}}{p_{2z} q_{2z}} \ .
 \label{Phase-space}
\end{eqnarray}
The pion propagators of eq.(\ref{Full-M-amp}) can be rewritten as
\begin{eqnarray}
p_1\cdot q_2&=&
  \frac{1}{2x} \left[ \mathbf{q}_{2\bot}^2 + x^2 m_{\pi}^2 \right]\label{Propagator1}\\
 p_2\cdot q_2&=&
  \frac{1}{2x'} \left[ (\mathbf{q}_{2\bot}-x' \mathbf{p}_{2\bot})^2+ x'^2 m_{\pi}^2 \right]
 \label{Propagator2}
\end{eqnarray}
with the parameters
\begin{eqnarray}
x&=&
  \frac{q_{2z}}{p_1} =\frac{\omega_2}{E_1}\label{x-fraction}\\
 x'&=&\frac{q_{2z}}{p_{2z}} =\frac{\omega_2}{E_2}=\frac{x}{1-x} \ .
 \label{xprim-fraction}
\end{eqnarray}

The matrix element in eq.(\ref{Full-M-amp}) can be further 
simplified. Since the polarisation vector $\bmath{\epsilon}_2$ lies in the plane
orthogonal to $\bmath{q}_2$,  the scalar products with the polarisation vector
can be  replaced by
 \begin{eqnarray}
\bmath{p}_1\cdot \bmath{\epsilon}_2&=&-\frac{1}{x}\,\bmath{q}_{2\bot}\cdot \bmath{\epsilon}_2 \\
\bmath{p}_2\cdot \bmath{\epsilon}_2&=&(\bmath{p}_{2\bot}-\frac{1}{x'}\,\bmath{q}_{2\bot}) \cdot \bmath{\epsilon}_2 \\
 (\bmath{p}_2 - \bmath{p}_1)\cdot \bmath{\epsilon}_2&=&\bmath{q}_{1\bot} \cdot \bmath{\epsilon}_2 \ ,
 \label{eps_products}
\end{eqnarray}
where perpendicular still means perpendicular to the incident pion direction. The new 
expression for the matrix element becomes more transparent
\begin{eqnarray}
{\cal M}&=&16\pi ie M_A  \left[  \bigg\{
	 \frac{E_2\mathbf{q}_{2\bot}}{\mathbf{q}_{2\bot}^2 + x^2 m_{\pi}^2 }
	   - \frac{E_1(\mathbf{q}_{2\bot}-x' \mathbf{p}_{2\bot})} 
	   { (\mathbf{q}_{2\bot}-x' \mathbf{p}_{2\bot})^2+ x'^2 m_{\pi}^2 }\bigg\}
	      {\mathcal F}_{rad}(\bmath{q}_{1\bot})\right.\nonumber \\
	    &&\left. \quad \quad+\frac{\lambda}{2}\frac{\omega_2\mathbf{q}_{1\bot}}{m_{\pi}^2}\,
	      {\mathcal F}_{pol}(\bmath{q}_{1\bot})\right]\cdot\bmath{\epsilon}_2 
	 \label{Simple-amp} 
\end{eqnarray}
and the summation over the photon polarisation directions embarrasingly trivial
\begin{eqnarray}
	\sum \left| {\cal M}\right|^2&=&64 (2\pi)^2 e^2 M_A^2 
	 \nonumber \\
	 &&\times \left| \left[ \frac{E_2\mathbf{q}_{2\bot}}{\mathbf{q}_{2\bot}^2 + x^2 m_{\pi}^2 }
	   - \frac{E_1(\mathbf{q}_{2\bot}-x' \mathbf{p}_{2\bot})} 
	   { (\mathbf{q}_{2\bot}-x' \mathbf{p}_{2\bot})^2+ x'^2 m_{\pi}^2 }\right]\right.
	      {\mathcal F}_{rad}(\bmath{q}_{1\bot})\nonumber \\
	    &&\left. \quad \quad+\frac{\lambda}{2}\frac{\omega_2\mathbf{q}_{1\bot}}{m_{\pi}^2}\,
	      {\mathcal F}_{pol}(\bmath{q}_{1\bot})\right|^2
	 \label{Amp-squared} 
\end{eqnarray}

We recall that $\bmath{q}_1$ is the momenum transfer to the nucleus and
 \begin{equation}
	\bmath{q}_{1}=\bmath{q}_{2}+\bmath{p}_{2}
\end{equation}
with $\bmath{p}_{2}$ the momentum of the final state pion and $\bmath{q}_{2}$
the momentum of the final state photon. In eq.(\ref{Amp-squared}) 
only the dependence on $\bmath{q}_{1\bot}$ is indicated. The longitudinal
component is fixed by the relation 
\begin{equation}
	- q_{1z}=\frac{1}{2p_1}[2p_2\cdot q_2+\bmath{q}_{1\bot}^2] \ .
	 \label{Min-mom-tr}
\end{equation}
For all practical purposes it can be replaced by its value in the forward
direction, $-q_{1z}=q_{min}$, with $q_{min}$ as given by eq.(\ref{min-mom}).
As a consequence, the denominator $\bmath{q}_1^2=\bmath{q}_{1\bot}^2+ q_{1z}^2$ 
of the Coulomb amplitude 
 never vanishes. However, the factor 
multiplying the form factor ${\mathcal F}_{rad}$ in eq.(\ref{Amp-squared})
vanishes in the forward direction, since there
 $\bmath{p}_1\cdot \bmath{\epsilon}_2 =\bmath{p}_2\cdot \bmath{\epsilon}_2=0$.

The cross section distribution  is dominated by the 
first term of eq.(\ref{Amp-squared}) which is multiplied by the 
form factor ${\mathcal F}_{rad}$. The second term is proportional to
the polarisability parameter $\lambda$ and the form factor  ${\mathcal F}_{pol}$
and is generally smaller. Since  
${\mathcal F}_{rad}$ is exactly the amplitude we encounter in elastic 
pion-nucleus scattering, this observation allows us to 
distinguish three regions, characterised by increasing values 
of the momentum transfer:
\begin{enumerate}
	\item {dominance of Coulomb bremsstrahlung}
	 	\item {competition between Coulomb and nuclear bremsstrahlung}
	 		\item {dominance of nuclear bremsstrahlung}
\end{enumerate}

In order to localise the proper Coulomb region we first localise
the overlap region where Coulomb and nuclear contributions are
of similar size. That happens when in the form factor ${\cal F}_{rad}$,
eq.(\ref{Decomp-C+N}), the Coulomb $f_C$ and  nuclear $f_N$ amplitudes
are equal in size. This, according to eq.(\ref{Forw-nucl-value}), is
roughly the case when  
\begin{equation}
	\frac{2Z\alpha}{\bmath{q}_{1\bot}^2}=\frac{1}{2}R_u^2 \ .
\end{equation}
Putting $Z=A/2$ as a further approximation gives
\begin{equation}
	\bmath{q}_{1\bot}^2\approx\frac{2\alpha}{r_0^2}A^{1/3}=
	0.5\cdot10^{-3}A^{1/3}\quad {\rm (GeV/c)}^2 \ .
\end{equation}
Consequently, in the transition region we have for a typical nucleus
$|t|\approx10^{-3}$ (GeV/c)$^2$; in the Coulomb-dominated region 
$|t|\approx10^{-4}$ (GeV/c)$^2$; and in the nuclear-dominated region 
$|t|\approx10^{-2}$ (GeV/c)$^2$. 

In the form factor ${\mathcal F}_{pol}$ the nuclear contribution
$f_{P}$ is substantially smaller than the corresponding $f_{N}$ in
${\mathcal F}_{rad}$ and would thus need larger momentum transfers 
to make itself noticed.
For Cu the ratio $|f_P(E,0)/f_N(E,0)|=0.12$ derived from the formulae
in the Appendix, supports this statement. 

There are  kinematic variables, beside those chosen above,
 that are of fundamental interest, {\it e.g.},
\begin{eqnarray}
	t&=&q_1^2\\
	s&=&(p_2+q_2)^2\\
	\Delta&=& (p_2-p_1)^2
\end{eqnarray}
which relate to properties of the underlying pion-Compton scattering
process. In fact, $t$ is the squared momentum transfer to the nucleus, but also
the squared mass of the virtual photon; $s$ is the square of the c.m.~energy
  and $\Delta$ the squared 
momentum transfer in the  pion-Compton  scattering process.
A simple calculation ends in the results
\begin{eqnarray}
	t&=&-\bmath{q}_{1\bot}^2-\left(\frac{\omega_2m_{\pi}}{2E_1E_2}\right)^2m_{\pi}^2 \\
	s&=&\frac{E_1}{E_2}m_{\pi}^2+\frac{1}{x'} (\mathbf{q}_{2\bot}-x' \mathbf{p}_{2\bot})^2\\
	\Delta&=& -\frac{E_1}{E_2}\left[\mathbf{p}_{2\bot}^2+x^2m_{\pi}^2\right] \ .
\end{eqnarray}
From the expression for $s$ we conclude that in the Coulomb region
the Compton scattering takes place at a lab energy of 
\[
\omega_{lab}=\half x'm_{\pi}\ .
\]
Similarly, from the expression for $\Delta$ we conclude that the scattering angle 
in the pion-Compton c.m.~system is $\theta_{cm}=90^\circ$, and in the lab system
\[
 \cos\theta_{lab} =-\frac{E_1-E_2}{E_1+E_2}=-\frac{x}{2-x}\ .
 \]
When $x'$ is large, which is an interesting region, then $\omega_{lab}$
 is also large, and one can legitimately ask where our basic 
amplitude for the Compton process breaks down.

Finally, there is the question of region of applicability of the
model described above. It is commonly stated that the radiation 
from the external legs
dominates as long as the propagator denominators, 
eqs  (\ref{Propagator1}) and (\ref{Propagator2}),   are of order
$m_{\pi}^2$ or smaller. Our formulae should therefore work 
well into the nuclear region. A quantification
of this  statement would mean estimating  diagrams where the 
photon is radiated from internal lines. We leave that investigation
to a future paper.
%
%
%
%%%%%%%%%%%%%%%%%%%%%%%%%
\newpage
\section{Under the spell of the Coulomb peak}

Pionic bremsstrahlung at very small momentum transfers, {\it i.e.} 
 in the Coulomb region,
 is of special interest
since it is there one can hope to extract the pion polarisability. 

In the Coulomb region the transverse momentum transfers are so small
they can safely be neglected in the propagators of 
eqs (\ref{Propagator1}) and (\ref{Propagator2}). 
Furthemore, as explained above,  in this region the  
Coulomb contributions dominate 
the form factors ${\mathcal F}_{rad}$ and ${\mathcal F}_{pol}$, 
and both may be replaced by $f_C(E,\bmath{q}_{\bot})/E$ from eq.(\ref{Point-cross-final}).
After some straightforward simplifications we get the cross section
distribution
\begin{equation}
\frac{\rd \sigma}{\rd^2q_{2\bot}  \rd^2p_{2\bot} \rd x}
= \frac{4Z^2\alpha^3}{\pi^2m_{\pi}^4}
 \Bigg[ \frac{\bmath{q}_{1\bot}^2}{(\bmath{q}_{1\bot}^2+q_{min}^2)^2}\Bigg]
  \Bigg[ \frac{1-x}{x^3} \left( 1+\frac{\lambda x^2}{2(1-x)}\right)^2\Bigg]
 \label{Coul-peak-cross}
 \end{equation}
 with $x=\omega_2/E_1$ from eq.(\ref{x-fraction}).
 
The factor inside the first pair of brackets is typical for Coulomb production.
It has the Coulomb propagator, with the minimum momentum transfer, but 
due to the numerator it vanishes
in the forward direction. This last property can readily be seen 
from expression (\ref{Coulomb_Born_lab}). In the forward direction, 
the pion momenta $\bmath{p}_1$ and $\bmath{p}_2$ are both parallel to the 
photon momentum $\bmath{q}_2$. Hence,
$\bmath{p}_1\cdot \bmath{\epsilon}_2 =\bmath{p}_2\cdot \bmath{\epsilon}_2=0$ and
  the ${\mathcal M}$-amplitude vanishes.
 
The factor inside the second pair of brackets contain the dependence on
the pion polarisability. Since from chiral-Lagrangian theory, eq.(\ref{lambda-def}),
we expect $\lambda\approx-0.01$, this dependence is very weak. Let us define 
$R(x)$ as
\begin{equation}
  R(x)=
   \left|1+\frac{\lambda x^2}{2(1-x)}\right|^2 \ .
\end{equation}
Suppose then that the bremsstrahlung photon carries half of the energy 
of the incident
pion. This gives $R(0.5)=0.99$, and a tiny  1\% effect from
the polarisability. To make a difference, the photon must take nearly
all the energy of the incident pion. As an example  $R(0.95)=0.78$,
a healthy 20\% effect. 
 But, increasing the $x-$value implies 
according to eq.(\ref{Coul-peak-cross}) at the same time diminishing
the cross section. Going from $x=0.5$ to $x=0.95$ means a reduction 
in cross section by a factor of 100. We conclude
that it is possible, but certainly very difficult,
to measure the pion polarisability in pion bremsstrahlung experiments.

Our result is at variance with that of Antipov  {\it et al.}\cite{5}.
In ref.\cite{12} a formula for the factor $R$ is given. It differs from
ours in that only the term linear in $ \lambda$ is kept, and it comes 
with a different factor. \\[2ex]

\noindent
{\bfseries Acknowledgements}. I would like to thank Barabara Badelek for 
drawing my attention to the 
importance of pionic bremsstrahlung.

%
%
%%%%%%%%%%%%%%%%%%%%%%
\newpage
\section{Appendix}
In order to estimate the  size of
the nuclear contributions we calculate the corresponding forward
amplitudes for uniform nuclear mass and charge distributions. 
A typical choice of  nuclear radius parameter is
\begin{equation}
	R_u=r_0 A^{1/3}
\end{equation}
with $r_0=1.1$ fm.
The normalised uniform density is then
\begin{equation}
	\rho(\bmath{r})=\frac{3}{4\pi R_u^3}\,\theta(R_u-r)\equiv\rho_0\theta(R_u-r)
\end{equation}
so that the mass and charge distributions become  $\rho_A(\bmath{r})=A\rho(\bmath{r})$ and  $\rho_Z(\bmath{r})=Z\rho(\bmath{r})$.

The forward elastic {\it nuclear amplitude}, void of
 Coulomb distortion, is according to eq.(\ref{fCN-new})
\begin{equation}
	f_N(E,0)= ik\int_0^{R_u}b\rd b\, \big(1-\exp[-\half\sigma' T_A(b)]\big)
	\label{Forw-nucl-value}
\end{equation}
with $T_A(b)=2A\rho_0R_u\sqrt{1-b^2/R_u^2}$. As always $\sigma'=\sigma(1-i\alpha)$, 
with $\sigma$ the pion-nucleon total cross section 
and $\alpha$ the phase of the forward elastic pion-nucleon scattering amplitude.
In terms of the parameter
\begin{equation}
	\xi= \frac{3\sigma'}{4\pi r_0^2} A^{1/3}
\end{equation}
we get
\begin{equation}
	f_N(E,0)= ikR_u^2\left[ \frac{1}{2}- \frac{1}{\xi^2} 
	+e^{-\xi}\left(\frac{1}{\xi} +\frac{1}{\xi^2}\right)\right] \ .
\end{equation}

The forward {\it polarisability amplitude}, void of
Coulomb distortion, is according to eq.(\ref{fCP-new})
\begin{equation}
	f_{P}(E,0) 
	   =-2 \pi \alpha E
    \int_0^{\infty}b\rd b\, \left[ \int_{0}^b b'\rd b' \,T_Z(b')\right]
     \left( 1- \exp[-\half \sigma' T_A(b)]\right)
\end{equation}
A straightforward integration gives 
\begin{equation}
	f_P(E,0)= -\alpha Z E R_u^2\left[ \frac{3}{10}- \frac{1}{\xi^2}
	 + \frac{24}{\xi^5}
	-e^{-\xi}\left(\frac{24}{\xi^5} + \frac{24}{\xi^4}+ \frac{12}{\xi^3}
	+\frac{3}{\xi^2} \right)\right]. 
\end{equation}
The scale factor  $\alpha Z$ in front makes $f_P$ smaller than $f_N$

If we take a cross section $\sigma=25$ mb, and $\alpha=0$, 
then the numerical parameter is $\xi=0.49A^{1/3}$.
%
%
% 
%
%%%%%%%%%%%%%%%%%%%%%%%%%%%%%%%%%%%%%%%% 

\end{document}